\documentclass[11pt,showpacs,preprint,preprintnumbers,nofootinbib,superscriptaddress,amsmath,amssymb]{revtex4}

\usepackage{graphicx}
\usepackage{dcolumn}
\usepackage{bm}
\usepackage{amssymb}
\usepackage{amsmath}
\usepackage{epsfig}    
\usepackage{color}
\usepackage{slashed}

\begin{document} 

\title{Constraints on Parameter Space from Perturbative Unitarity\\
 in Models with Three Scalar Doublets}

%
\author{Stefano Moretti}
\email{S.Moretti@soton.ac.uk}
\affiliation{School of Physics and Astronomy, University of Southampton, Southampton, SO17 1BJ, United Kingdom}
\author{Kei Yagyu}
\email{K.Yagyu@soton.ac.uk}
\affiliation{School of Physics and Astronomy, University of Southampton, Southampton, SO17 1BJ, United Kingdom}

\begin{abstract}

We calculate an $s$-wave amplitude matrix for all the possible 2--to--2 body scalar boson elastic scatterings 
in models with three scalar doublets, including contributions from the longitudinal component of weak gauge bosons 
via the Equivalence Theorem Approximation. Specifically, we concentrate on the two cases with two[one] active plus one[two] inert doublet fields, 
referred to as I(1+2)HDM[I(2+1)HDM], under CP conservation. 
We obtain three analytically irreducible sub-matrices with the $3\times 3$ form and eighteen eigenvalues for the amplitude matrix as an independent set, 
where the same formula can be applied to both models. 
By requiring a perturbative unitarity condition, we can constrain the magnitude of quartic coupling constants in the Higgs potential. 
This constraint, in particular in the I(1+2)HDM, 
can be translated into a bound on masses of extra active scalar bosons. 
Furthermore, 
when Standard Model-like Higgs boson couplings with weak gauge bosons 
are deviated from the Standard Model predictions, the unitarity condition provides an upper limit on the masses.  
We find that  
stronger upper bounds on the masses of the active CP-even and CP-odd Higgs bosons are obtained under the 
constraints from the unitarity and vacuum stability conditions, as well as the electroweak $S$, $T$ and $U$ parameters,
as compared to those 
in 2-Higgs Doublet Models with a softly-broken $Z_2$ symmetry.  

\end{abstract}

\maketitle

\section{Introduction}

After the completion of the operations for LHC Run-I, the most important legacy for the whole of particle physics is
the existence of a Higgs boson with a mass of about 126 GeV~\cite{Higgs}. Furthermore, as
a result of combined data from various Higgs search channels, it was established  that 
the observed Higgs boson has properties which are consistent with those of 
the Standard Model (SM) Higgs state~\cite{SM_Higgs}. 

However, it is natural to be left with a key question.
Namely, what is the actual number of scalar doublet fields present in Nature?
Although the Higgs sector in the SM is constructed of only one isospin doublet field, 
that is not supported by a fundamental principle. 
In fact, it is well known that multi-doublet models can be considered so that the Higgs boson phenomenology
is consistent with current LHC data by arranging additional parameters suitably, and 
they are often introduced new physics models beyond the SM. 
 
For example, Supersymmetry-extended models require at least two doublets in their Higgs sector 
for gauge anomaly cancellation and to construct the Yukawa Lagrangian~\cite{susy}. 
A second doublet field is introduced also in many radiative neutrino mass models~\cite{radseesaw}. 
The Inert Doublet Model (IDM)~\cite{IDM}, aka a 2-Higgs Doublet Model (2HDM) with one active and one inert 
doublet, has been proposed to explain the existence of Dark Matter (DM),
by depriving a Vacuum Expectation Value (VEV) of the second doublet field and making it 
non-interacting with SM fields except for the (thereby active) Higgs boson itself and the gauge states\footnote{This can be simply realized by 
imposing an additional unbroken discrete $Z_2$ symmetry into the Higgs sector, 
where all the SM particles [the second doublet] are assigned to be even [odd] under the $Z_2$ symmetry.  
The particles with the odd parity are produced in pairs and the lightest one amongst these is stable, hence
a DM candidate.}.  
Furthermore, an additional source of CP-violation is obtained in a Higgs sector with multiple doublets 
which is required to realize a successful scenario for  Electro-Weak (EW) baryogenesis~\cite{ewbg}. 

Properties of an enlarged Higgs sector clearly depend on the new physics scenario embedding it, so that  
the determination of the true Higgs boson dynamics is extremely important to extract information on the new physics 
model. In particular,  if a multi-doublet structure is chosen for the Higgs sector, 
what is a mass scale of the additional active and/or inert (pseudo)scalar bosons? 
How can we extract information on the total number of doublet fields?
Needless to say, 
the direct way to probe a multi-doublet structure 
is discovering extra (pseudo)scalar bosons at collider experiments. 
For example, even in the simplest case of 2HDMs, 
there appear a pair of singly-charged ($H^\pm$), a CP-odd ($A$) and a CP-even ($H$) Higgs boson in addition to the SM-like Higgs boson ($h$). 
Clearly, the most important property to find these extra scalar bosons is their masses. 

Although, in general, these masses are taken as free parameters,  
we may be able to constrain their corresponding parameter space by taking into account theoretical constraints: chiefly, 
perturbative unitarity and vacuum stability. 
In 2HDMs, the unitarity  and vacuum stability bounds have been discussed in Refs.~\cite{PU_THDM1,PU_THDM2} and in Ref.~\cite{VS_THDM}, respectively. 
By using them, dimensionless parameters in the scalar potential can be constrained, 
and it can be translated into limits on the mass parameters for the extra (pseudo)scalar bosons. 
In particular, when there is a non-zero mixing between $h$ and $H$ in 2HDMs, 
the exact decoupling limit of the extra scalar boson can no longer be taken, because of the unitarity constraints.  
Therefore, one can obtain an upper limit on the mass of the extra scalar boson~\cite{fingerprint}. The argument can
be generalized to multi-Higgs doublet scenarios. 

The effects of a non-zero mixing can appear as deviations in the couplings of $h$  
with gauge bosons ($hVV$) and/or fermions ($hf\bar{f}$) from the SM predictions. Hence, the very fact that
the latter can be accessed experimentally enables one to extract information on the underlying (pseudo)scalar dynamics
above and beyond the mass scale directly probed at the collider concerned. Take the $hVV$ couplings for
example. They are presently (i.e., after LHC Run-I) constrained to be SM-like at the level of 10\% or so.
LHC Run-II (13 TeV) at standard integrated luminosity (300 fb$^{-1}$) would improve this precision by about a factor of 2.
Their expected accuracy will then be about 2\%~\cite{hVV} at the High-Luminosity (HL) LHC~\cite{HLLHC} 
(with collision energy of 14 TeV and integrated luminosity of 3000 fb$^{-1}$). 
Furthermore, the accuracy is expected to be about 0.4\%~\cite{hVV} at the International Linear Collider (ILC) with collision energy of 500 GeV and integrated luminosity of 500 fb$^{-1}$ (ILC500).
Therefore, if a deviation in $hVV$ couplings is discovered at future colliders, whenever this will occur,  
it will at the same time suggest the possibility of an enlarged Higgs sector and, if so, 
enable one to place an upper limit on the mass of extra (pseudo)scalar bosons.  

In this paper, we investigate constraints on dimensionless parameters in the scalar potential from the bounds 
imposed by perturbative unitarity and vacuum stability 
in 3-Higgs Doublet Models (3HDMs). The phenomenological 
 importance of such extended Higgs scenarios, amongst many possible others,
has been emphasized lately in the context of the latest LHC results and DM searches. 
For example, 3HDMs may shed light on the flavour problem, namely the unknown origin and nature of the three families of
quarks and leptons, including neutrinos, and their pattern of masses, mixings and CP
violation: it is possible that the three families of quarks and leptons could be described
by the same symmetries that describe the three Higgs doublets \cite{15,Hartmann:2014ppa}.  
Such a family symmetry could be spontaneously broken along with the EW symmetry,
although some remnant subgroup could survive, thereby stabilizing a possible scalar
DM candidate while for certain symmetries it is possible to find a VEV
alignment that respects the original symmetry of the potential which will then
be responsible for the stabilization of the DM candidate \cite{13}.

Although there are several types of 3HDMs depending on symmetries imposed in the scalar potential 
\cite{14}--\cite{3hdm_s3}, having in view the aforementioned phenomenological requirements and
a need for a simple and predictive formulation, recently, $Z_2$ symmetric 3HDMs with one or two 
inert doublets have been investigated and found viable from both a DM and LHC
perspective \cite{3HDM_pheno,3HDM_DM}. Furthermore, other than explaining current data, 
new regions of DM relic density and areas of parameter space yielding totally new LHC signals
(with respect to, e.g., 2HDMs) open up \cite{3HDM_pheno,3HDM_DM,cb}. 

Herein, we  focus on the I(1+2)HDM and I(2+1)HDM, where the former[latter] model contains 
two[one] active doublets and one[two] inert doublets. 
We calculate the $s$-wave amplitude matrix for all possible 2--to--2 body scalar boson elastic scatterings 
in the high energy limit in both models, including contributions from 
the SM gauge bosons as the Nambu-Goldstone (NG) boson states due to the Equivalence Theorem Approximation (ETA)\footnote{In Ref.~\cite{3hdm_s3}, 
the similar approach has been applied to a 3HDM with a non-Abelian discrete $S_3$ symmetry to obtain eigenvalues for the $s$ wave amplitude matrix. 
Because the symmetry imposed in the Higgs potential in our paper is different from that in Ref.~\cite{3hdm_s3}, 
derived eigenvalues are also different. We have confirmed that our formula for the eigenvalues coincides with those presented in Ref.~\cite{3hdm_s3}
by taking appropriate replacements of the coupling constant in the potential. Detailed explanations are given in the end of Sec.~III-B. }.
We then examine the allowed parameter space from the unitarity and vacuum stability (re-derived herein
and found to be compliant with those in Refs.~\cite{3HDM_pheno})
bounds in addition to the constraints from the  $S$, $T$ and $U$ parameters. 
Then, as intimated, we proceed to apply the unitarity bounds to obtain an upper limit on the masses of $A$ and $H$ in the I(1+2)HDM in the case with
a non-zero mixing between $h$ and $H$, and study how the limit can be changed as compared to that in 2HDMs. 
We find that a stronger upper bound is obtained in the I(1+2)HDM than that in 2HDMs depending on the masses of the inert (pseudo)scalar bosons.  
Therefore, we can exclude the I(1+2)HDM when the masses of $A$ or $H$ will be found to be larger than the given upper limit for a fixed amount of the mixing 
between $h$ and $H$. 
As for the I(2+1)HDM, since unitarity here is perfectly realized by the SM-like
Higgs state $h$, we study what constraints are induced on the various parameters in the potential 
involving inert states) in order not to spoil the perturbative behavior of the only active scalar in this scenario.  

This paper is organized as follows. 
In Sec.~II, we give the scalar potential and mass formulae for the scalar bosons. 
In Sec.~III, constraints from vacuum stability, perturbative unitarity and the $S$, $T$ and $U$ parameters are discussed. 
In Sec.~IV, we delineate the allowed parameter regions from the constraints given in the preceding section. Finally, 
conclusions are given in Sec.~V.

\begin{table}[t]
\begin{center}
{\renewcommand\arraystretch{1}
\begin{tabular}{c|ccc|ccc}\hline\hline
&\multicolumn{3}{c|}{I(1+2)HDM} & \multicolumn{3}{c}{I(2+1)HDM}\\\hline 
Doublet & $\varphi_0$ & $\varphi_1$ & $\varphi_2$ & $\varphi_0$ & $\varphi_1$ & $\varphi_2$ \\\hline 
$Z_2$&$-$&$+$&$+$&+&$-$&$-$    \\\hline 
$\tilde{Z}_2$&+&+&$-$&+&$+$&$-$   \\\hline \hline
\end{tabular}}
\caption{Charge assignment under the $Z_2\times \tilde{Z}_2$ symmetry in the I(1+2)HDM and I(2+1)HDM. }
\label{particle}
\end{center}
\end{table}

\section{Three Higgs Doublet Models}

\subsection{The scalar potential}

Among various classes of 3HDMs~\cite{3HDM1}, we discuss the one  which satisfies the following requirements:
\begin{description}
\item{(i)} it contains a SM-like Higgs boson; 
\item{(ii)} it contains a unique candidate of DM in a given mass spectrum; 
\item{(iii)} it does not have Flavour-Changing Neutral Current (FCNCs) at the tree level. 
\end{description}
In order to satisfy the requirements of (i) and (ii), we need at least one active and one inert doublet field.
Thus, the (pseudo)scalar sector of our 3HDMs must be composed of one inert plus two active doublet fields; i.e., I(1+2)HDM or  
two inert plus one active doublet fields; i.e., I(2+1)HDM.
To guarantee a doublet field being inert and the stability of a DM candidate, 
we require the corresponding VEV to be zero and
impose an unbroken discrete $Z_2$ symmetry onto the whole doublet sector. 
A DM candidate can then be obtained as the lightest neutral (pseudo)scalar component of the inert doublet field with the odd parity under $Z_2$ 
(the SM plus the Higgs/active (pseudo)scalar states being even). As a consequence,
we also obtain that the inert (pseudo)scalar states can only appear in pairs in each vertex in the Lagrangian, so that 
they cannot couple to fermions directly. 
In the I(1+2)HDM, FCNC processes via neutral scalar boson mediation appear at the tree level similarly 
to the generic 2HDM case, because 
of the existence of two doublet fields with the same quantum numbers. 
The simplest way to avoid such FCNCs is imposing another discrete ${Z}_2$ symmetry~\cite{GW} which can be softly-broken in general (henceforth denoted as  $\tilde{Z}_2$). 
Here, we consider both the I(1+2)HM and I(2+1)HDM with the two discrete symmetries $Z_2$ and $\tilde{Z}_2$\footnote{In the I(2+1)HDM, 
there is no FCNC at the tree level even when we do not impose the $\tilde{Z}_2$ symmetry. However, by imposing the $\tilde{Z}_2$ symmetry throughout,  the structure of the Higgs potential can be much simpler than that in the case without it, thereby rendering possible a semi-analytical treatment of unitarity violating processes. 
In this connection, it turns out that we can deal with the two models uniformly as seen in the scalar potential (given below in Eq.~(\ref{pot})). We thus consider the $Z_2\times \tilde{Z}_2$ symmetric potential in both 3HDMs. }. 

To write down the scalar potential in the 3HDMs, 
let $\varphi_i$ $(i=0,...2)$ be isosipin doublet scalar fields, 
where the assignment of $Z_2\times \tilde{Z}_2$ charge is listed in Table~\ref{particle}. 
The most general scalar potential is then given in the same form in both the I(1+2)HDM and I(2+1)HDM by 
\begin{align}
V(\varphi_0,\varphi_1,\varphi_2) &= \sum_{i=0,...2}\mu_i^2 \varphi_i^\dagger \varphi_i + (\mu_{12}^2 \varphi_1^\dagger \varphi_2 + \text{h.c.}) \notag\\
 & +\frac{1}{2}\sum_{i=0,...2}\lambda_i (\varphi_i^\dagger \varphi_i)^2 +\lambda_3(\varphi_1^\dagger \varphi_1)(\varphi_2^\dagger \varphi_2)
+\lambda_4|\varphi_1^\dagger \varphi_2|^2 +\frac{1}{2}[\lambda_5(\varphi_1^\dagger \varphi_2)^2 + \text{h.c.} ] \notag\\
&+ \rho_1(\varphi_1^\dagger\varphi_1)(\varphi_0^\dagger\varphi_0)
 +\rho_2|\varphi_1^\dagger \varphi_0|^2
+\frac{1}{2}[\rho_3(\varphi_1^\dagger \varphi_0)^2 + \text{h.c.} ]\notag\\
&+ \sigma_1(\varphi_2^\dagger\varphi_2)(\varphi_0^\dagger\varphi_0)
 +\sigma_2|\varphi_2^\dagger \varphi_0|^2
+\frac{1}{2}[\sigma_3(\varphi_2^\dagger \varphi_0)^2 + \text{h.c.} ], \label{pot}
\end{align}
where $\mu_{12}^2$, $\lambda_5$, $\rho_3$ and $\sigma_3$ are complex parameters in general. 
Throughout the paper, we take these parameters to be real for simplicity, thereby avoiding explicit CP violation.
We also neglect the possibility of complex VEVs, thus also removing spontaneous CP violation. Hence, 
the (pseudo)scalar fields can be parameterised as 
\begin{align}
\varphi_i = \left[
\begin{array}{cc}
H_i^+ \\ 
\frac{1}{\sqrt{2}}(H_i+v_i+iA_i)
\end{array}\right],~~(i=0,...2), 
\end{align}
where $v_i$ are the VEVs of $\varphi_i$ with the sum rule $\sum_{i}v_i^2=v^2\simeq (246$ GeV$)^2$. 
In the I(2+1)HDM, $v_1=v_2=0$ and $v_0=v$ while in the I(1+2)HDM, $v_1,~v_2\neq 0$ and $v_0=0$. 

The Yukawa Lagrangian in the I(2+1)HDM is given in the same form as that in the SM, where 
$\varphi_0$ gives all the SM fermion masses. In contrast,  in the I(1+2)HDM, 
the structure of the Yukawa Lagrangian depends on the $\tilde{Z}_2$ charge assignments for the SM fermions. 
In general, there are four independent choices of the assignment~\cite{4types,4types2} and these are the so-called Type-I, Type-II, Type-X and Type-Y~\cite{typeX}. 
Various constraints from flavour \cite{stal} and collider physics \cite{typeX,Guedes,Chiang-Yagyu,fingerprint} strongly depend
on this choice. However,  
the following discussion does not, though we will make sure to avoid phenomenologically dangerous regions of
parameter space.

\subsection{Mass formulae in the I(1+2)HDM}

In the I(1+2)HDM, the inert sector is composed of one doublet. Therefore, 
the CP-even, CP-odd and pair of charged scalar states in the doublet field $\varphi_0$ correspond to the mass 
eigenstates. 
To avoid a confusion between the inert and active scalar states, we represent 
the inert states $H_0^\pm$, $A_0$ and $H_0$ as $\eta^\pm$, $\eta_A^{}$ and $\eta_H^{}$, respectively. 
Their masses are calculated as 
\begin{align}
m_{\eta^\pm}^2 & = \mu_0^2 + \frac{v^2}{2}\left[\rho_1  \cos^2\beta + \sigma_1 \sin^2\beta \right], \\ 
m_{\eta_H^{}}^2 & = \mu_0^2 + \frac{v^2}{2}\left[(\rho_1 + \rho_2 + \rho_3)\cos^2\beta+(\sigma_1 + \sigma_2 + \sigma_3)\sin^2\beta \right], \\
m_{\eta_A^{}}^2 & = \mu_0^2 + \frac{v^2}{2}\left[(\rho_1 + \rho_2 - \rho_3)\cos^2\beta+(\sigma_1 + \sigma_2 - \sigma_3)\sin^2\beta \right]. 
\end{align}

The mass formulae for the active sector are completely the same as those in a generic 2HDM, 
so that we can directly apply the same mass formulae to the I(1+2)HDM. 
The mass eigenstates for the active scalar bosons are thus  given as: 
\begin{align}
&\begin{pmatrix}
H_1^\pm \\
H_2^\pm 
\end{pmatrix}
= R(\beta)
\begin{pmatrix}
G^\pm \\
H^\pm 
\end{pmatrix},~
\begin{pmatrix}
A_1^{} \\
A_2^{}
\end{pmatrix}
= R(\beta)
\begin{pmatrix}
G^0 \\
A 
\end{pmatrix},
\begin{pmatrix}
H_1^{} \\
H_2^{}
\end{pmatrix}
= R(\alpha)
\begin{pmatrix}
H \\
h 
\end{pmatrix},\notag\\
&\text{with}~~R(\theta) = 
\begin{pmatrix}
\cos\theta & -\sin\theta \\
\sin\theta & \cos\theta
\end{pmatrix}, 
\end{align}
where $G^\pm$ and $G^0$ are the NG bosons which are absorbed by a longitudinal component of 
the $W^\pm$ and $Z$ bosons, respectively, and 
$h$ is the SM-like Higgs boson. 
The mixing angle $\beta$ is expressed by the ratio of the VEVs as $\tan\beta = v_2/v_1$. 

The squared masses of $H^\pm$ and $A$ are then calculated as 
\begin{align}
m_{H^\pm}^2=M^2-\frac{v^2}{2}(\lambda_4+\lambda_5),\quad m_A^2&=M^2-v^2\lambda_5, \label{massch}
\end{align}
where 
\begin{align}
M^2 = -\frac{\mu_{12}^2}{\sin\beta\cos\beta}. 
\end{align}
The mass matrix for the neutral CP-even (scalar) states is  expressed in terms of $\beta$ and 
the usual mixing angle $\alpha$ as 
\begin{align}
&m_H^2=\cos^2(\alpha-\beta)M_{11}^2+\sin^2(\alpha-\beta)M_{22}^2+\sin2(\alpha-\beta)M_{12}^2, \label{mass2} \\
&m_h^2=\sin^2(\alpha-\beta)M_{11}^2+\cos^2(\alpha-\beta)M_{22}^2-\sin2(\alpha-\beta)M_{12}^2,\\
&\tan 2(\alpha-\beta)=\frac{2M_{12}^2}{M_{11}^2-M_{22}^2}.  \label{tan2a}
\end{align}
where the matrix elements are 
\begin{align}
M_{11}^2&=v^2(\lambda_1\cos^4\beta+\lambda_2\sin^4\beta)+\frac{v^2}{2}(\lambda_3+\lambda_4+\lambda_5)\sin^22\beta,\notag\\
M_{22}^2&=M^2+v^2\sin^2\beta\cos^2\beta\left[\lambda_1+\lambda_2-2(\lambda_3+\lambda_4+\lambda_5)\right],\notag\\
M_{12}^2&=\frac{v^2}{2}\sin2\beta(-\lambda_1\cos^2\beta+\lambda_2\sin^2\beta)+\frac{v^2}{2}\sin2\beta\cos2\beta(\lambda_3+\lambda_4+\lambda_5). \label{mateven}
\end{align}

One of the most important variables in the active sector of the I(1+2)HDM, or equivalently in a 2HDM, is 
$\sin(\beta-\alpha)$.
This describes deviations in the SM-like Higgs boson $h$ couplings to gauge bosons and fermions. Namely, 
the ratios of the $hVV$ ($V=W^\pm,~Z$) and $hf\bar{f}$ couplings in the I(1+2)HDM to those in the SM  are given by
\begin{align}
\frac{g_{hVV}^{\text{I(1+2)HDM}}}{g_{hVV}^{\text{SM}}} = \sin(\beta-\alpha), \quad
\frac{g_{hff}^{\text{I(1+2)HDM}}}{g_{hff}^{\text{SM}}} = \sin(\beta-\alpha)+\xi_f \cos(\beta-\alpha), \label{couplings}
\end{align} 
where the
$\xi_f$ factors are $\cot\beta$ or $-\tan\beta$ depending on the type of fermion (i.e.,  up-type and down-type quarks and charged leptons) and 
the type of Yukawa interactions (i.e., the aforementioned 2HDM Types)\footnote{The $\xi_f$ factor corresponds to 
$2T_3^f\xi_A^f$ with $T_3^f$ being the third component of the isospin, where $\xi_A^f$ is given in Table~II of Ref.~\cite{typeX}.  }.  
The point here is that when we take the limit of $\sin(\beta-\alpha)\to 1$, both the $hVV$ and $hf\bar{f}$ couplings 
become the same as those in the SM. We thus call this limit the SM-like configuration. Furthermore, by looking at Eq.~(\ref{mateven}), 
we can see that only $M_{22}^2$ contains the term proportional to $M^2$, while $M_{11}^2$ and $M_{12}^2$ are proportional to $v^2$.   
Thus, when we take the limit of $M^2\to \infty$, the mixing angle given in Eq.~(\ref{tan2a}) becomes $\tan 2(\alpha-\beta)\to 0$, 
which also gives $m_H^2 \to M_{22}^2$ and $m_h^2 \to M_{11}^2$ by choosing $\alpha-\beta\to -\pi/2$.
Therefore, the extra Higgs bosons $H^\pm$, $A$ and $H$ are decoupled due to the $M^2$ term in their mass formulae, so that 
we can call this limit the decoupling limit~\cite{Gunion-Haber}. 

From the above discussion, we can reach an important conclusion regarding the nature of the Higgs sector in this model. 
The SM-like limit is naturally achieved when the decoupling limit is taken. 
In other words, the decoupling limit cannot be realized when we consider the case of a deviation from the SM-like limit, because 
we need a strong cancellation between $M_{11}^2$ and $M_{22}^2$ in the denominator of Eq.~(\ref{tan2a}),
in order to have $\tan 2(\alpha-\beta)\neq 0$. 
This would force the dimensionless coupling constants $\lambda_i$ in the Higgs potential to acquire quite large values which must be excluded from the view point of ensuring perturbativity of the model. 
This in turn implies that there is an upper limit on the masses of the extra Higgs bosons when we consider the case of $\sin(\beta-\alpha)\neq 1$. As mentioned previously, one of the 
aims of this paper is indeed deriving an upper limit for these in the I(1+2)HDM 
by using bounds from vacuum stability and perturbative unitarity, which we will discuss in the next section. 

\subsection{Mass formulae in the I(2+1)HDM}
 
Next, we present the mass formulae in the I(2+1)HDM. 
In this model, $\varphi_0$ corresponds to the SM Higgs doublet, so that 
$H_0^\pm$ and $A_0$ correspond to the NG bosons $G^\pm$ and $G^0$, respectively, 
while $H_0$ is the SM-like Higgs boson $h$. 

The mass of $h$ is given just like the SM Higgs boson as 
\begin{align}
m_h^2 = 2\lambda_0 v^2. \label{mh}
\end{align}
In contrast, the inert sector is composed of two doublets, 
so that the masses for singly-charged, CP-odd and CP-even states are, respectively, given in the 
$2\times 2$ matrix form. 
The mass matrix for the singly-charged scalar states is evaluated in the basis of ($H^\pm_1$, $H_2^\pm$) as 
\begin{align}
\mathcal{M}_C^2 = \left( 
\begin{array}{cc}
\mu_1^2 +\frac{v^2}{2}\rho_1 &  \mu_{12}^2  \\
\mu_{12}^2 & \mu_{2}^2 + \frac{v^2}{2}\sigma_1
\end{array}\right), 
\end{align}
and that for the CP-odd and CP-even states are given in the basis of ($H_1^{}$,$H_2^{}$) and ($A_1^{}$,$A_2^{}$), respectively, 
\begin{align}
\mathcal{M}_{A}^2 = \mathcal{M}_C^2 + \frac{v^2}{2}\left(
\begin{array}{cc}
 \rho_2 - \rho_3 & 0 \\
 0   &   \sigma_2 - \sigma_3
\end{array}\right), ~~
\mathcal{M}_{H}^2 = \mathcal{M}_C^2+\frac{v^2}{2}\left( 
\begin{array}{cc}
\rho_2 + \rho_3  & 0 \\
0   & \sigma_2 +  \sigma_3. 
\end{array}\right). 
\end{align}
The mass eigenstates can be defined by introducing three mixing angles, as follows:
\begin{align}
\begin{pmatrix}
H_1^\pm \\
H_2^\pm 
\end{pmatrix}
= R(\theta_C)
\begin{pmatrix}
\eta_1^\pm \\
\eta_2^\pm 
\end{pmatrix},~
\begin{pmatrix}
A_1^{} \\
A_2^{}
\end{pmatrix}
= R(\theta_A)
\begin{pmatrix}
\eta_{A_1}^{} \\
\eta_{A_2}^{} 
\end{pmatrix},~
\begin{pmatrix}
H_1^{} \\
H_2^{}
\end{pmatrix}
= R(\theta_H)
\begin{pmatrix}
\eta_{H_1}^{} \\
\eta_{H_2}^{} 
\end{pmatrix}.  
\end{align}
The squared mass eigenvalues for the singly-charged states ($\eta_1^\pm$, $\eta_2^\pm$) are expressed in terms of the elements of the mass matrix as 
\begin{align}
m_{\eta_1^\pm}^2 &= \cos^2\theta_C (\mathcal{M}_C^2)_{11} +\sin^2\theta_C (\mathcal{M}_C^2)_{22}+\sin2\theta_C (\mathcal{M}_C^2)_{12}, \\ 
m_{\eta_2^\pm}^2 &= \sin^2\theta_C (\mathcal{M}_C^2)_{11} +\cos^2\theta_C (\mathcal{M}_C^2)_{22}-\sin2\theta_C (\mathcal{M}_C^2)_{12}, 
\end{align}
while those for the CP-odd and CP-even states ($\eta_{A_1}^{}$, $\eta_{A_2}^{}$) and ($\eta_{H_1}^{}$,  
 $\eta_{H_2}^{}$), respectively, are obtained by replacing $(\theta_C,{\cal M}^2_C)$ with $(\theta_A,{\cal M}^2_A)$ and $(\theta_H,{\cal M}^2_H)$. 
Thus, the three mixing angles are given by 
\begin{align}
\tan2\theta_X = \frac{2(\mathcal{M}_X^2)_{12}}{(\mathcal{M}_X^2)_{11}-(\mathcal{M}_X^2)_{22}},~~\text{for}~~X=C,A,H. 
\end{align}

In summary, the inert sector of the I(2+1)HDM can be described by nine parameters: i.e., six masses $m_{\eta_i^\pm}$, $m_{\eta_{A_i}^{}}$ and $m_{\eta_{H_i}^{}}$ for $i=1,2$ and 
three mixing angles $\theta_X$ (for $X=C,A,H$). 
These nine physical parameters are expressed in terms of the nine parameters in the potential 
of Eq. (\ref{pot}): i.e., $\mu_1^2$, $\mu_2^2$, $\mu_{12}^2$, 
$\rho_i$ and $\sigma_i$ ($i=1,2,3$). 

\section{Constraints on the parameter space}

In this section, we discuss constraints on the parameter space in the I(1+2)HDM and the I(2+1)HDM. 
As theoretical constraints, we discuss  bounds from vacuum stability and perturbative unitarity. 
As experimental constraints, other than ensuring sampling the 3HDM parameter spaces concerned away from experimental
limits extracted via direct Higgs searches at LEP, Tevatron and LHC 
we also crucially take into account the so-called $S$, $T$ and $U$ parameters proposed by Peskin 
and Takeuchi~\cite{stu} in the high precision EW fits to LEP data \cite{Baak:2012kk}. 

For the discussion of vacuum stability and unitarity, there is no difference between in the I(1+2)HDM and I(2+1)HDM. 
In contrast, for the $S$, $T$ and $U$ parameters, different formulae are presented here in each of two models. 

\subsection{Vacuum stability}

The Higgs potential should be bounded from below in any direction of the scalar boson space. 
The sufficient condition to guarantee such a positivity of the potential is given by  the following requirements:
\begin{align}
& \lambda_0 > 0,\quad \lambda_1 > 0,\quad \lambda_2 > 0, \\
& \sqrt{\lambda_1\lambda_2} + \lambda_3 + \text{MIN}(0,~\lambda_4+\lambda_5,~\lambda_4-\lambda_5)>0, \\
& \sqrt{\lambda_0\lambda_1} + \rho_1 + \text{MIN}(0,~\rho_2+\rho_3,~\rho_2-\rho_3)>0, \label{vs2} \\
& \sqrt{\lambda_0\lambda_2} + \sigma_1 + \text{MIN}(0,~\sigma_2+\sigma_3,~\sigma_2-\sigma_3)>0. \label{vs3} 
\end{align}

\subsection{Unitarity}

Constraints on the scalar self-coupling constants in the potential are extracted
from the requirement of $S$-matrix unitarity for all the elastic scatterings of two body scalar boson states.  
This idea to constrain a scalar self-coupling has been proposed in Ref.~\cite{lqt} for the Higgs sector in the SM and 
it has been applied to constrain the mass of the SM Higgs boson. It can however be generalized to multi-doublet models,
as well known \cite{HHG}. 
The requirement of the $S$-matrix unitarity is translated into a relationship for the $J$th partial wave amplitude $a_J$. 
In the high energy limit, it can be expressed as 
\begin{align}
\text{Im}(a_J) = |a_J|^2. 
\end{align}
This implies that $a_J$ should be on a circle with radius and center of $1/2$ and $(0,1/2)$ in the complex plane, respectively. 
From the above relation, we can require the following condition for the tree level amplitude of $a_J$ as~\cite{HHG} 
\begin{align}
|\text{Re}(a_J)| < \frac{1}{2}. \label{lqtbound}
\end{align}
In the high energy limit, each element of $a_J$ for all possible $S_1S_2 \to S_3S_4$ processes, 
where the $S_i$'s represent all (pseudo)scalar bosons in the model, is simply given by the (pseudo)scalar four point interaction. 
In that case, only the $s$-wave amplitude $(J=0)$ can contribute to the scattering process. 
We thus apply the inequality given in Eq.~(\ref{lqtbound}) to the case of $J=0$.  

There are 30, $36~(=18\times 2)$ and $12~(=6\times 2)$ channels for electromagnetically neutral, singly-charged and doubly-charged states, respectively, 
where half of the singly- and doubly-charged states correspond to the charge conjugated states of the remaining half.  
Each channel classified by the difference in the Electro-Magnetic (EM) charge is orthogonal to the others, so that  
we can separately consider each of the channels. 
In order to distinguish these classes, we define $a_0^N$, $a_0^C$ and $a_0^D$ as the $s$-wave amplitude for the neutral, singly-charged and doubly-charged states, respectively. 

In addition to the classification in terms of the EM charge, 
it is also helpful to distinguish the states by using the $Z_2\times \tilde{Z}_2$ quantum numbers of the 
intervening external states.  Because we are now focusing on the (pseudo)scalar boson quartic interactions only, 
the softly-broken $\tilde{Z}_2$ parity can be considered as the exact quantum number. 
Under $Z_2\times \tilde{Z}_2$, four types of (pseudo)scalar pairings are allowed: i.e., 
$(Z_2,\tilde{Z}_2)=(+,+),~(+,-)$, $(-,+)$ and $(-,-)$, where each state does not mix with any other one. 
We thus further decompose the matrices for the $s$-wave amplitude into four sub-matrices as follows:
\begin{align}
a_0^N = \text{diag}\left(a_{++}^N,~a_{+-}^N,~a_{-+}^N,~a_{--}^N\right),  \\
a_0^{C} = \text{diag}\left(a_{++}^C,~a_{+-}^C,~a_{-+}^C,~a_{--}^C\right),  \\
a_0^{D} = \text{diag}\left(a_{++}^{D},~a_{+-}^{D},~a_{-+}^{D},~a_{--}^{D}\right). 
\end{align}
For the singly- and doubly-charged states, we focus on the 18[6] channels with the positive charge, because 
the remained negative charged states are connected by the charge conjugation as explained.

\subsubsection{Neutral channels}

We first consider the neutral channels.
There are 12 pairings, $H_i^+H_i^-$, $A_iA_i/\sqrt{2}$, $H_iH_i/\sqrt{2}$ and $A_iH_i$ ($i=0,\,1,\,2$) with $(Z_2,\tilde{Z}_2)=(+,+)$.  
The 6 states with $(+,-)$ are $H_1^+H_2^-$, $H_1^-H_2^+$, $A_1A_2$, $H_1H_2$, $A_1H_2$ and $H_1A_2$.  
The other states with $(-,+)$ and ($-,-$) are respectively given 
by replacing the subscript of the ($+,-$) states by $X_1Y_2\to X_0Y_1$ and $X_1Y_2\to X_0Y_2$.
In fact, each sub-matrix in $a_0^N$ can  be made even simpler by choosing the 
following basis for the $(+,+)$ states:
\begin{align}
|\Psi_{++}^N\rangle_{i+1} &= \frac{1}{\sqrt{2}}\left(H_i^+H_i^- + \frac{1}{\sqrt{2}}A_iA_i + \frac{1}{\sqrt{2}}H_iH_i\right), \\
|\Psi_{++}^N\rangle_{i+4} &= \frac{1}{\sqrt{2}}\left(H_i^+H_i^- - \frac{1}{\sqrt{2}}A_iA_i - \frac{1}{\sqrt{2}}H_iH_i\right), \\
|\Psi_{++}^N\rangle_{i+7} &= \frac{1}{\sqrt{2}}\left(A_iA_i - H_iH_i\right), \\
|\Psi_{++}^N\rangle_{i+10}  &=  A_iH_i, ~~\text{with}~~(i=0,...2). 
\end{align}
For the $(+,-)$ states, we can choose 
\begin{align}
|\Psi_{+-}^N\rangle_1 &= \frac{1}{2}\left(H_1^+H_2^- + H_1^-H_2^+ +A_1A_2 + H_1H_2  \right) , \\
|\Psi_{+-}^N\rangle_2 &= \frac{1}{2}\left(-H_1^+H_2^- - H_1^-H_2^+ + A_1A_2 + H_1H_2  \right) , \\
|\Psi_{+-}^N\rangle_3 &= \frac{1}{2}\left(iH_1^+H_2^- -i H_1^-H_2^+ -A_1H_2 + H_1A_2  \right) , \\
|\Psi_{+-}^N\rangle_4 &= \frac{1}{2}\left(-iH_1^+H_2^- +i H_1^-H_2^+ -A_1H_2 + A_1H_2  \right) , \\
|\Psi_{+-}^N\rangle_5 &= \frac{i}{\sqrt{2}}\left(A_1H_2 + H_1A_2   \right), \\
|\Psi_{+-}^N\rangle_6 &= \frac{1}{\sqrt{2}}\left(-A_1A_2 + H_1H_2   \right). 
\end{align} 
Similarly, the basis for the $(-,+)$ and $(-,-)$ states denoted as $|\Psi_{-+}^N\rangle_j$ and $|\Psi_{--}^N\rangle_j$ ($j=1 ... 6$) 
can be chosen by replacing the subscript in the above $(+,-)$ states by $X_1Y_2\to X_0Y_1$ and $X_1Y_2\to X_0Y_2$, respectively. 
In these basis, each sub-matrix in $a_0^N$ can be  block-diagonal form with a $3\times 3$ sub-matrix as the largest one,
\begin{align}
a_{++}^N &=\frac{1}{16\pi}\text{ diag}\left[ 
(N_{++}^{3\times 3})_1,
(N_{++}^{3\times 3})_2,
(N_{++}^{3\times 3})_3,
(N_{++}^{3\times 3})_4\right], \\
a_{+-}^N &= \frac{1}{16\pi}\text{diag}\left[n_{+-}^1,n_{+-}^2,n_{+-}^3,n_{+-}^4,n_{+-}^5,n_{+-}^6 \right], \\
a_{-+}^N &= \frac{1}{16\pi}\text{diag}\left[n_{-+}^1,n_{-+}^2,n_{-+}^3,n_{-+}^4,n_{-+}^5,n_{-+}^6 \right], \\
a_{--}^N &= \frac{1}{16\pi}\text{diag}\left[n_{--}^1,n_{--}^2,n_{--}^3,n_{--}^4,n_{--}^5,n_{--}^6 \right], 
\end{align}
where 
\begin{align}
(N_{++}^{3\times 3})_1 & = 
\begin{pmatrix} 
3\lambda_0 &2\rho_1 + \rho_2 & 2\sigma_1 + \sigma_2  \\
2\rho_1 + \rho_2 & 3\lambda_1 & 2\lambda_3 + \lambda_4\\
 2\sigma_1 + \sigma_2 & 2\lambda_3 + \lambda_4 & 3\lambda_2 
\end{pmatrix}, \\
(N_{++}^{3\times 3})_2 & =
\begin{pmatrix} 
\lambda_0 &\rho_2 & \sigma_2  \\
\rho_2 & \lambda_1 &  \lambda_4\\
\sigma_2 & \lambda_4 & \lambda_2 
\end{pmatrix}, \\
(N_{++}^{3\times 3})_3 & = (N_{++}^{3\times 3})_4= 
\begin{pmatrix} 
\lambda_0 &\rho_3 & \sigma_3  \\
\rho_3 & \lambda_1 &  \lambda_5\\
\sigma_3 & \lambda_5 & \lambda_2 
\end{pmatrix},\\ 
n_{+-}^1 & = \lambda_3 + 2\lambda_4 + 3\lambda_5, \\
n_{+-}^2 & = \lambda_3 +\lambda_5, \\
n_{+-}^3 & = \lambda_3 -\lambda_5, \\
n_{+-}^4 & = \lambda_3 + 2\lambda_4 - 3\lambda_5, \\
n_{+-}^5 & = n_{+-}^6 =  \lambda_3 + \lambda_4 , \\
n_{-+}^1 & = \rho_1 + 2\rho_2 + 3\rho_3, \\
n_{-+}^2 & = \rho_1 +\rho_3, \\
n_{-+}^3 & = \rho_1 -\rho_3, \\
n_{-+}^4 & = \rho_1 + 2\rho_2 - 3\rho_3, \\
n_{-+}^5 & = n_{-+}^6 =  \rho_1 + \rho_2 , \\
n_{--}^1 & = \sigma_1 + 2\sigma_2 + 3\sigma_3, \\
n_{--}^2 & = \sigma_1 +\sigma_3, \\
n_{--}^3 & = \sigma_1 -\sigma_3, \\
n_{--}^4 & = \sigma_1 + 2\sigma_2 - 3\sigma_3, \\
n_{--}^5 & = n_{--}^6 =  \sigma_1 + \sigma_2. 
\end{align}

\subsubsection{Singly-charged channels}

Second, we consider the 18 positive singly-charged channels. 
The sub-matrix $a_{++}^C$ has the form of $6\times 6$, while the sub-matrices $a_{+-}^C,~a_{-+}^C$ and $a_{--}^C$ have 
the form of $4\times 4$. 
The basis of $a_{++}^C$ can be expressed as $(H_i^+ A_i,H_i^+ H_i)$ ($i=0,...2$), and that of 
$a_{+-}^C$ can be expressed by $(H_1^+ A_2,H_1^+ H_2,H_2^+ A_1,H_1^+ H_2)$. 
The basis for $a_{-+}^C$ and $a_{--}^C$ are respectively given by replacing the subscript of the basis for $a_{+-}^C$ by 
$(1,2)\to (0,1)$ and $(1,2)\to (0,2)$. 

By choosing the following basis, we obtain a further simplified form of each sub-matrix:
\begin{align}
|\Psi_{++}^C \rangle_{i} &= \frac{1}{\sqrt{2}}\left(iH_i^+ A_i + H_i^+ H_i \right), \\
|\Psi_{++}^C \rangle_{i+3} &= \frac{1}{\sqrt{2}}\left(iH_i^+ A_i - H_i^+ H_i \right),  ~~\text{with}~~(i=0,...2), \\
|\Psi_{+-}^C \rangle_1 &= \frac{1}{2}\left(iH_1^+ A_2 + H_1^+ H_2 + iA_1 H_2^+ + A_1 H_2^+ \right), \\
|\Psi_{+-}^C \rangle_2 &= \frac{1}{2}\left(iH_1^+ A_2 + H_1^+ H_2 - iA_1 H_2^+ - A_1 H_2^+ \right), \\
|\Psi_{+-}^C \rangle_3 &= \frac{1}{2}\left(iH_1^+ A_2 - H_1^+ H_2 + iA_1 H_2^+ - A_1 H_2^+ \right), \\
|\Psi_{+-}^C \rangle_4 &= \frac{1}{2}\left(iH_1^+ A_2- H_1^+ H_2- iA_1 H_2^+ + A_1 H_2^+ \right). 
\end{align}
The basis for the $(-,+)$ and $(-,-)$ states denoted as $|\Psi_{-+}^C\rangle_k$ and $|\Psi_{--}^C\rangle_k$ ($k=1,...4$) 
can be chosen by replacing the subscript in the above $(+,-)$ states by $X_1Y_2\to X_0Y_1$ and $X_1Y_2\to X_0Y_2$, respectively. 
In this basis, each sub-matrix in $a_0^C$ can be  block-diagonal form as
\begin{align}
a_{++}^C &= \frac{1}{16\pi}\text{diag}\left[(C_{++}^{3\times 3})_1, (C_{++}^{3\times 3})_2 \right], \\
a_{+-}^C &= \frac{1}{16\pi}\text{diag}\left[c_{+-}^1,c_{+-}^2,c_{+-}^3,c_{+-}^4 \right], \\
a_{-+}^C &= \frac{1}{16\pi}\text{diag}\left[c_{-+}^1,c_{-+}^2,c_{-+}^3,c_{-+}^4 \right], \\
a_{--}^C &= \frac{1}{16\pi}\text{diag}\left[c_{--}^1,c_{--}^2,c_{--}^3,c_{--}^4 \right], 
\end{align}
where 
\begin{align}
&(C_{++}^{3\times 3})_1 = (N_{++}^{3\times 3})_2, \quad (C_{++}^{3\times 3})_2 = (N_{++}^{3\times 3})_3, \\ 
&c_{+-}^1 = n_{+-}^2, \\
&c_{+-}^2 = n_{+-}^3, \\
&c_{+-}^3 = n_{+-}^5, \\
&c_{+-}^4 = \lambda_3 - \lambda_4,\\
&c_{-+}^1 = n_{-+}^2, \\
&c_{-+}^2 = n_{-+}^3, \\
&c_{-+}^3 = n_{-+}^5, \\
&c_{-+}^4 = \rho_1 - \rho_2,\\
&c_{--}^1 = n_{--}^2, \\
&c_{--}^2 = n_{--}^3, \\
&c_{--}^3 = n_{--}^5, \\
&c_{--}^4 = \sigma_1 - \sigma_2. 
\end{align}
We note that $c_{+-}^{4}$, $c_{-+}^{4}$ and $c_{--}^{4}$ give 
the independent eigenvalues against those in the  neutral states.  

\subsubsection{Doubly-charged channels}

Finally, we consider the 6 positive doubly-charged channels. 
The sub-matrix $a_{++}^D$ has the form of $3\times 3$ in the basis of $(H_i^+ H_i^+)/\sqrt{2}$ ($i=0,...2$). 
Further, $a_{+-}^D,~a_{-+}^D$ and $a_{--}^D$ have the $1\times 1$ form for the $H_1^+H_2^+$, $H_0^+H_1^+$ and $H_0^+H_2^+$ states, respectively. 
In the doubly-charged channels, the above basis gives already the simplest structure of the matrix of the $s$-wave amplitude,  written as  
\begin{align}
a_{++}^D &= \frac{1}{16\pi}(D_{++}^{3\times 3}), \\
a_{+-}^D &= \frac{1}{16\pi}d_{+-}, \\
a_{-+}^D &= \frac{1}{16\pi}d_{-+}, \\
a_{+-}^D &= \frac{1}{16\pi}d_{--}. 
\end{align}
They result in a $3\times 3$ matrix with eigenvalues given as 
\begin{align}
&(D_{++}^{3\times 3})= (N_{++}^{3\times 3})_3, \\
&d_{+-} =n_{+-}^5,\quad  d_{-+} = n_{-+}^5,\quad  d_{--}= n_{--}^5. 
\end{align}

\subsubsection{Summary}

Consequently, we get 3 independent sub-matrices and 18 eigenvalues for $a_0$. 
After we rename these, they are finally expressed by 
\begin{align}
X_1 &= 
\begin{pmatrix} 
3\lambda_0 &2\rho_1 + \rho_2 & 2\sigma_1 + \sigma_2  \\
2\rho_1 + \rho_2 & 3\lambda_1 & 2\lambda_3 + \lambda_4\\
 2\sigma_1 + \sigma_2 & 2\lambda_3 + \lambda_4 & 3\lambda_2 
\end{pmatrix}, ~
X_2  =
\begin{pmatrix} 
\lambda_0 &\rho_2 & \sigma_2  \\
\rho_2 & \lambda_1 &  \lambda_4\\
\sigma_2 & \lambda_4 & \lambda_2 
\end{pmatrix},~
X_3 = 
\begin{pmatrix} 
\lambda_0 &\rho_3 & \sigma_3  \\
\rho_3 & \lambda_1 &  \lambda_5\\
\sigma_3 & \lambda_5 & \lambda_2 
\end{pmatrix}, \label{3times3}  \\ 
y_1^\pm & = \lambda_3 + 2\lambda_4 \pm 3\lambda_5, \label{y1}\\
y_2^\pm & = \rho_1 + 2\rho_2 \pm 3\rho_3, \\
y_3^\pm & = \sigma_1 + 2\sigma_2 \pm 3\sigma_3, \\
y_4^\pm & = \lambda_3 \pm \lambda_5, \\
y_5^\pm & = \rho_1 \pm \rho_3, \\
y_6^\pm & = \sigma_1 \pm \sigma_3, \\
y_7^\pm & =  \lambda_3 \pm \lambda_4 , \\
y_8^\pm &=  \rho_1 \pm \rho_2 , \\
y_9^\pm & =  \sigma_1 \pm \sigma_2.  \label{y9}
\end{align}
The condition in Eq.~(\ref{lqtbound}) is expressed in terms of the above variables as 
\begin{align}
|x_i|     & < 8\pi,~~(i=1,...9) , \\
|y_j^\pm| & < 8\pi,~~(j=1,...9) , 
\end{align}
where $x_i$ are the eigenvalues of $X_1$, $X_2$ and $X_3$. 

We note that all the eigenvalues of the matrices given in Eq.~(\ref{3times3}) and those in Eqs.~(\ref{y1})-(\ref{y9})
coincide with those given in Ref.~\cite{3hdm_s3}, in which the $s$ wave amplitude matrix has been derived in the $S_3$ symmetric version of the 3HDM, 
by taking the following replacement of the coupling constants 
\begin{align}
\lambda_1 &\to  2 (\lambda_1 + \lambda_3), \quad
\lambda_2 \to  2 (\lambda_1 + \lambda_3),\quad
\lambda_0 \to  2 \lambda_8, \notag\\ 
\lambda_3 &\to  2 (\lambda_1 - \lambda_3),\quad
\lambda_4 \to  2 (-\lambda_2 + \lambda_3),\quad
\lambda_5 \to  2 (\lambda_2 + \lambda_3), \notag\\ 
\rho_1 &\to  \lambda_5,\quad
\rho_2 \to  \lambda_6,\quad
\rho_3 \to  2 \lambda_7, \notag\\ 
\sigma_1 &\to  \lambda_5,\quad
\sigma_2 \to  \lambda_6,\quad
\sigma_3 \to  2 \lambda_7, 
\end{align}
with taking $\lambda_4$ in Ref.~\cite{3hdm_s3} to be zero. 

\subsection{$S$, $T$ and $U$ parameters}

We now discuss the $S$, $T$ and $U$  parameters  in the 
context of our 3HDMs. 
The differences between the $S$, $T$ and $U$ parameters  ($S_{\text{3HDM}}$, $T_{\text{3HDM}}$ and $U_{\text{3HDM}}$) 
and those in SM ($S_{\text{SM}}$, $T_{\text{SM}}$ and $U_{\text{SM}}$), are defined by  
\begin{align}
\Delta S = S_{\text{3HDM}} - S_{\text{SM}}, \notag\\
\Delta T = T_{\text{3HDM}} - T_{\text{SM}}, \notag\\
\Delta U = U_{\text{3HDM}} - U_{\text{SM}}. 
\end{align}
Throughout this subsection, we use the shortened notations $s_X^{}\equiv \sin X$ and $c_X^{}\equiv \cos X$. 

In the I(1+2)HDM, 
the differences are composed of two parts: i.e., 
the contribution from active scalar boson loops and that from inert scalar boson loops. 
The former one is the same as in the 2HDM while the latter one 
is the same as in the IDM, which is composed of one active and one inert doublet fields. 
Therefore, $\Delta S$, $\Delta T$ and $\Delta U$ can be just obtained by summing the two contributions,
active ($A$) and inert ($I$), as 
\begin{align}
\Delta S[\text{I(1+2)HDM}] = \Delta S_{\text{A}} + \Delta S_{\text{I}}, \notag\\
\Delta T[\text{I(1+2)HDM}] = \Delta T_{\text{A}} + \Delta T_{\text{I}}, \notag\\
\Delta U[\text{I(1+2)HDM}] = \Delta U_{\text{A}} + \Delta U_{\text{I}}. 
\end{align}
The $S$, $T$ and $U$ parameters in the 2HDM and IDM have been calculated in Ref.~\cite{stu_2hdm} and 
in Ref.~\cite{IDM}, respectively. 
Each of the differences is obtained as follows:  
\begin{align}
\Delta S_{\text{A}}& = \frac{1}{4\pi }\Bigg\{  s_{\beta-\alpha}^2F'(m_Z^2;m_H,m_A)-F'(m_Z^2;m_{H^\pm},m_{H^\pm}) \notag\\
& +c_{\beta-\alpha}^2\Big[F'(m_Z^2;m_h,m_A)+F'(m_Z^2;m_H,m_Z)-F'(m_Z^2;m_h,m_Z)\Big]\notag\\
& +4m_Z^2c_{\beta-\alpha}^2\Big[G'(m_Z^2;m_H,m_Z)-G'(m_Z^2;m_h,m_Z)\Big]\Bigg\}, \\
\Delta T_{\text{A}}& = \frac{1}{16\pi^2 \alpha_{\text{em}} v^2}
\Bigg\{ 
F(0;m_{H^\pm},m_A)
+s_{\beta-\alpha}^2[F(0;m_{H^\pm},m_H)-F(0;m_A,m_H)]\notag\\
& +c_{\beta-\alpha}^2\Big[F(0;m_{H^\pm},m_h)+F(0;m_H,m_W)+F(0;m_h,m_Z)\notag\\
&\quad\quad\quad -F(0;m_h,m_W)-F(0;m_A,m_h)-F(0;m_H,m_Z)\notag\\
&\quad\quad\quad + 4G(0;m_H,m_W)+4G(0;m_h,m_Z)-4G(0;m_h,m_W)-4G(0;m_H,m_Z)\Big]  \Bigg\}, \\
\Delta U_{\text{A}}& = \frac{1}{4\pi}
\Bigg\{ 
F'(m_W^2;m_{H^\pm},m_A)-F'(m_Z^2;m_{H^\pm},m_{H^\pm})\notag\\
&\quad\quad+s_{\beta-\alpha}^2[F'(m_W^2;m_{H^\pm},m_H)-F'(m_Z^2;m_A,m_H)]\notag\\
&\quad\quad +c_{\beta-\alpha}^2\Big[F'(m_W^2;m_{H^\pm},m_h)+F'(m_W^2;m_W,m_H)-F'(m_W^2;m_W,m_h)\Big]  \notag\\
&\quad\quad -c_{\beta-\alpha}^2\Big[F'(m_Z^2;m_A,m_h)+F'(m_Z^2;m_Z,m_H)-F'(m_Z^2;m_Z,m_h)\Big]  \notag\\
&\quad\quad +4m_W^2c_{\beta-\alpha}^2\Big[G'(m_W^2;m_H,m_W)-G'(m_W^2;m_h,m_W)\Big]\notag\\
&\quad\quad -4m_Z^2c_{\beta-\alpha}^2\Big[G'(m_Z^2;m_H,m_Z)-G'(m_Z^2;m_h,m_Z)\Big] \Bigg\}, \\
\Delta S_{\text{I}}&=\frac{1}{4\pi}
\Big[F'(m_Z^2;m_{\eta_H^{}},m_{\eta_A^{}}) - F'(m_Z^2;m_{\eta^\pm},m_{\eta^\pm})\Big],\\
\Delta T_\text{I}
&=
\frac{1}{16\pi^2 \alpha_{{\rm em}} v^2} 
\Big[F(0;m_{\eta^\pm},m_{\eta_A^{}}) 
+ F(0;m_{\eta^\pm},m_{\eta_H^{}})-F(0;m_{\eta_A^{}},m_{\eta_H^{}})  \Big],\\
\Delta U_\text{I}&=\frac{1}{4\pi}
\Big[F'(m_W^2;m_{\eta^\pm},m_{\eta_H^{}}) + F'(m_W^2;m_{\eta^\pm},m_{\eta_A^{}})\notag\\
&-F'(m_Z^2;m_{\eta^\pm},m_{\eta^\pm}) -F'(m_Z^2;m_{\eta_H},m_{\eta_A})   \Big],
\end{align}
where $F'(m_V^2;m_1,m_2)=[F(m_V^2;m_1,m_2)-F(0;m_1,m_2)]/m_V^2$ and $G'(m_V^2;m_1,m_2)=[G(m_V^2;m_1,m_2)-G(0;m_1,m_2)]/m_V^2$. 
The loop functions are given by 
\begin{align}
F(p^2;m_1,m_2)&=\int_0^1 dx \Big[(2x-1)(m_1^2-m_2^2)+(2x-1)^2p^2\Big]\ln \Delta_B, \\
F(0;m_1,m_2)&=\frac{1}{2}(m_1^2+m_2^2)+\frac{2m_1^2m_2^2}{m_1^2-m_2^2}\ln\frac{m_2}{m_1}, \\
G(p^2;m_1,m_2)&= \int_0^1 dx  \ln \Delta_B , \\
G(0;m_1,m_2)&=\ln(m_1m_2)-\frac{m_1^2+m_2^2}{m_1^2-m_2^2}\ln\frac{m_2}{m_1}-1, \\
\Delta_B&=  xm_1^2+(1-x)m_2^2-x(1-x)p^2. 
\end{align}
We note that the functions $F$ and $G$ given in the above are invariant by interchanging the second and the third arguments; i.e., 
$F(p^2;m_1,m_2)=F(p^2;m_2,m_1)$ and  $G(p^2;m_1,m_2)=G(p^2;m_2,m_1)$. 
We also note that $\Delta S_{\text{I}}$, $\Delta T_{\text{I}}$ and $\Delta U_{\text{I}}$ are obtained by 
taking $\sin(\beta-\alpha)=1$ and making the replacement $(m_{H^\pm},m_A,m_H)\to (m_{\eta^\pm}$, $m_{\eta_A},m_{\eta_H})$ in 
$\Delta S_{\text{A}}$, $\Delta T_{\text{A}}$ and $\Delta U_{\text{A}}$.

In the I(2+1)HDM, the differences purely come from the inert (pseudo)scalar boson loops. 
The analytic expressions are given  by 
\begin{align}
\Delta S=\frac{1}{4\pi}
&\Bigg\{- F'(m_Z^2;m_{\eta_1^\pm},m_{\eta_1^\pm})
 - F'(m_Z^2;m_{\eta_2^\pm},m_{\eta_2^\pm})\notag\\
&+c_{\theta_A-\theta_H}^{}\left[F'(0;m_{\eta_{A_1}^{}},m_{\eta_{H_1}^{}})+ F'(0;m_{\eta_{A_2}^{}},m_{\eta_{H_2}^{}})\right]\notag\\
&+s_{\theta_A-\theta_H}^{}\left[F'(0;m_{\eta_{A_1}^{}},m_{\eta_{H_2}^{}}) - F'(0;m_{\eta_{A_2}^{}},m_{\eta_{H_1}^{}})\right]
\Bigg\},\\
\Delta T
=
\frac{1}{16\pi^2 \alpha_{{\rm em}} v^2} 
&\Bigg\{
c_{\theta_C-\theta_A}^{}\left[F(0;m_{\eta_1^\pm},m_{\eta_{A_1}^{}}) + F(0;m_{\eta_2^\pm},m_{\eta_{A_2}^{}})\right]\notag\\
&+s_{\theta_C-\theta_A}^{}\left[F(0;m_{\eta_1^\pm},m_{\eta_{A_2}^{}}) - F(0;m_{\eta_2^\pm},m_{\eta_{A_1}^{}})\right]\notag\\
&+c_{\theta_C-\theta_H}^{}\left[F(0;m_{\eta_1^\pm},m_{\eta_{H_1}^{}}) + F(0;m_{\eta_2^\pm},m_{\eta_{H_2}^{}})\right]\notag\\
&+s_{\theta_C-\theta_H}^{}\left[F(0;m_{\eta_1^\pm},m_{\eta_{H_2}^{}}) - F(0;m_{\eta_2^\pm},m_{\eta_{H_1}^{}})\right]\notag\\
&-c_{\theta_A-\theta_H}^{}\left[F(0;m_{\eta_{A_1}^{}},m_{\eta_{H_1}^{}}) + F(0;m_{\eta_{A_2}^{}},m_{\eta_{H_2}^{}})\right]\notag\\
&-s_{\theta_A-\theta_H}^{}\left[F(0;m_{\eta_{A_1}^{}},m_{\eta_{H_2}^{}}) - F(0;m_{\eta_{A_2}^{}},m_{\eta_{H_1}^{}})\right]  \Bigg\},\\
\Delta U =\frac{1}{4\pi}
&\Bigg\{- F'(m_Z^2;m_{\eta_1^\pm},m_{\eta_1^\pm})
 - F'(m_Z^2;m_{\eta_2^\pm},m_{\eta_2^\pm})\notag\\
&+c_{\theta_C-\theta_H}^{}\left[F'(m_W^2;m_{\eta_1^\pm},m_{\eta_{H_1}^{}}) + F'(m_W^2;m_{\eta_2^\pm},m_{\eta_{H_2}^{}})\right]\notag\\
& +s_{\theta_C-\theta_H}^{}\left[F'(m_W^2;m_{\eta_1^\pm},m_{\eta_{H_2}^{}}) - F'(m_W^2;m_{\eta_2^\pm},m_{\eta_{H_1}^{}})\right]\notag\\
&+c_{\theta_C-\theta_A}^{}\left[F'(m_W^2;m_{\eta_1^\pm},m_{\eta_{A_1}^{}}) + F'(m_W^2;m_{\eta_2^\pm},m_{\eta_{A_2}^{}})\right]\notag\\
& +s_{\theta_C-\theta_A}^{}\left[F'(m_W^2;m_{\eta_1^\pm},m_{\eta_{A_2}^{}}) - F'(m_W^2;m_{\eta_2^\pm},m_{\eta_{A_1}^{}})\right]\notag\\
&-c_{\theta_A-\theta_H}^{}\left[F'(m_Z^2;m_{\eta_{A_1}^{}},m_{\eta_{H_1}^{}}) + F'(m_Z^2;m_{\eta_{A_2}^{}},m_{\eta_{H_2}^{}})\right]\notag\\
& -s_{\theta_A-\theta_H}^{}\left[F'(m_Z^2;m_{\eta_{A_1}^{}},m_{\eta_{A_2}^{}}) - F'(m_Z^2;m_{\eta_{A_2}^{}},m_{\eta_{A_1}^{}})\right]\Bigg\}. 
\end{align}

The deviations in the $S$ and $T$ parameters  from the SM predictions under
$m_h=126$ GeV and $U=0$ are given by~\cite{Baak:2012kk}
\begin{align}
\Delta S =0.05\pm 0.09,\quad  \Delta T = 0.08\pm 0.07,
\label{STallowed}
\end{align}
where the correlation factor between $\Delta S$ and $\Delta T$ is +0.91. 
We use the above experimental values for the numerical analysis in the next section.

\section{Numerical Analysis}

\begin{figure}[t]
\begin{center}
\includegraphics[scale=0.3]{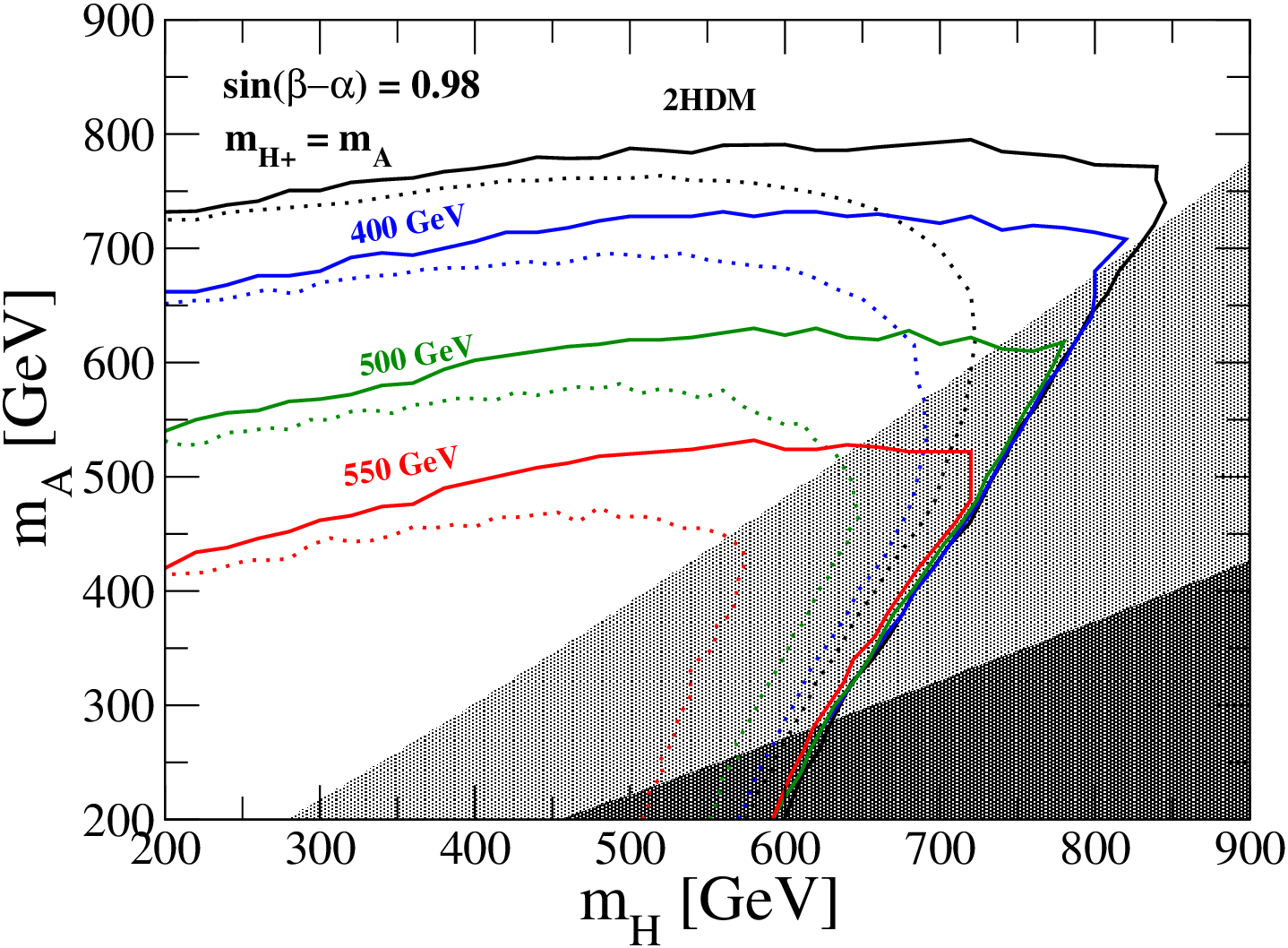} \hspace{3mm}
\includegraphics[scale=0.3]{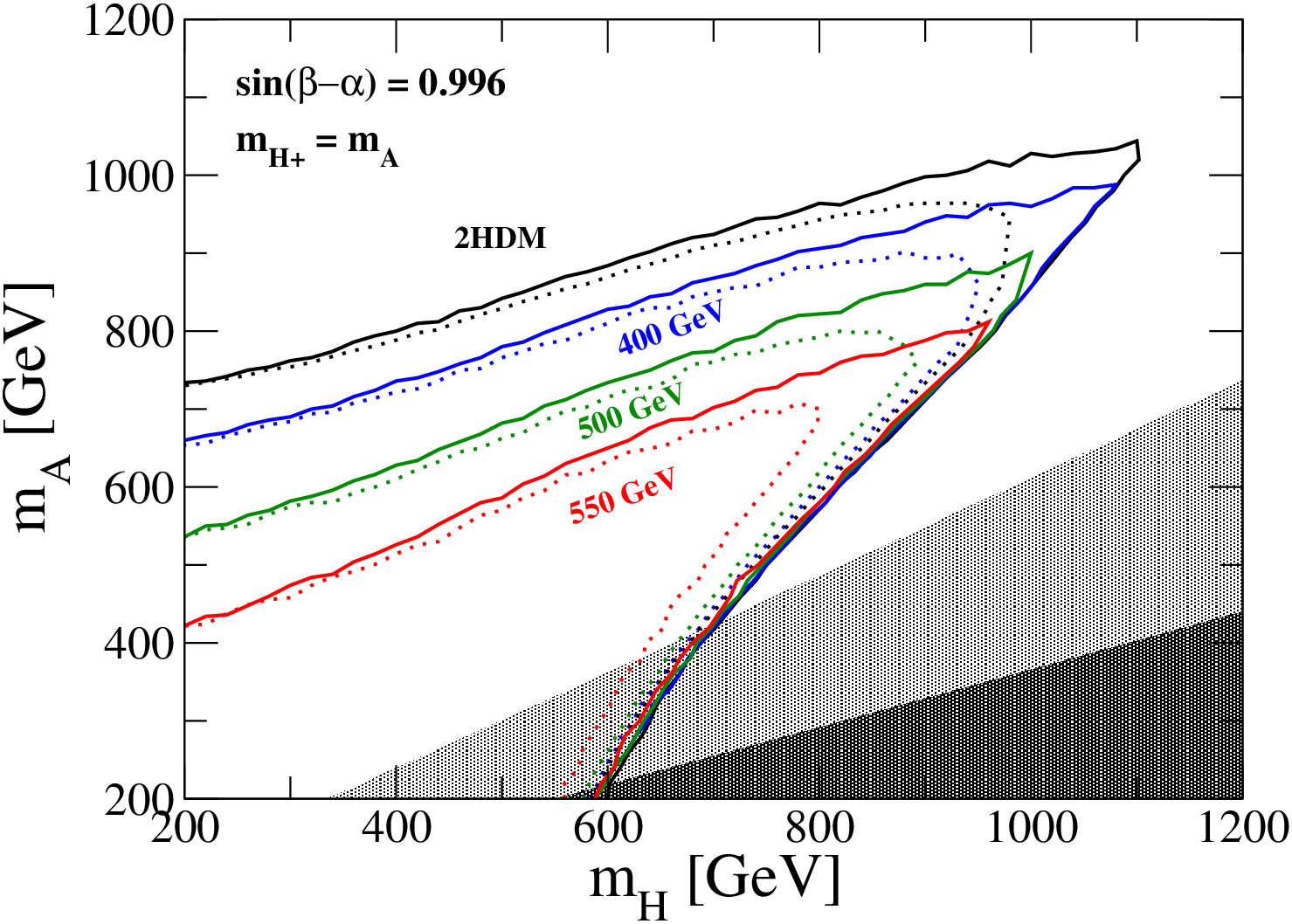}
\caption{
Constraints on the  I(1+2)HDM parameter space expressed in the 
the $m_H$-$m_A$ plane induced by unitarity, vacuum stability and the $S$ and $T$ parameters in the 
case of $m_{H^\pm}=m_A$, $m_{\eta^\pm}=m_{\eta_H}$,  $m_{\eta_A}=63$ GeV  and $\sin(\beta-\alpha)=0.98$ (left panel) 
and 0.996 (right panel).
We take $m_{\eta^\pm}=400$, 500 and 550 GeV for the blue, green and red contours. 
We also show the result in the 2HDM as the black contour for comparison. 
The value of $\tan\beta$ is fixed to be 1 in the dotted contours while it is scanned over the range of $1\leq \tan\beta \leq 30$ in the solid contours. 
For all the plots, we scan the value of $M^2$ in the range of $M^2=m_H^2 \pm 1$ TeV$^2$. 
The outside regions from each contour are excluded by unitarity and vacuum stability. 
The light and dark shaded regions are also excluded by $S$ and $T$ parameters in the 2HDM and  I(1+2)HDM, respectively. 
}
\label{const1}
\end{center}
\vspace{5mm}
\begin{center}
\includegraphics[scale=0.3]{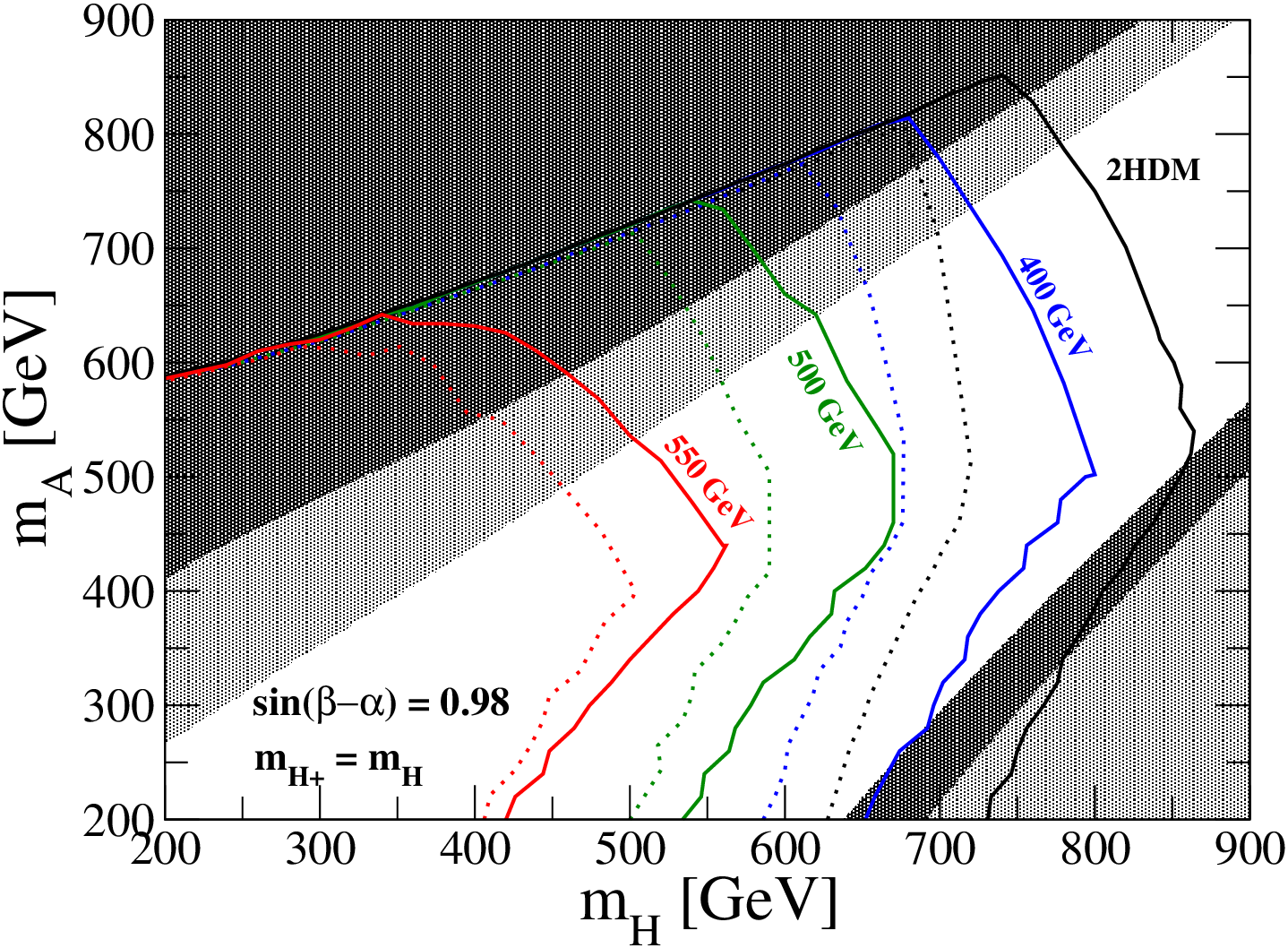} \hspace{3mm}
\includegraphics[scale=0.3]{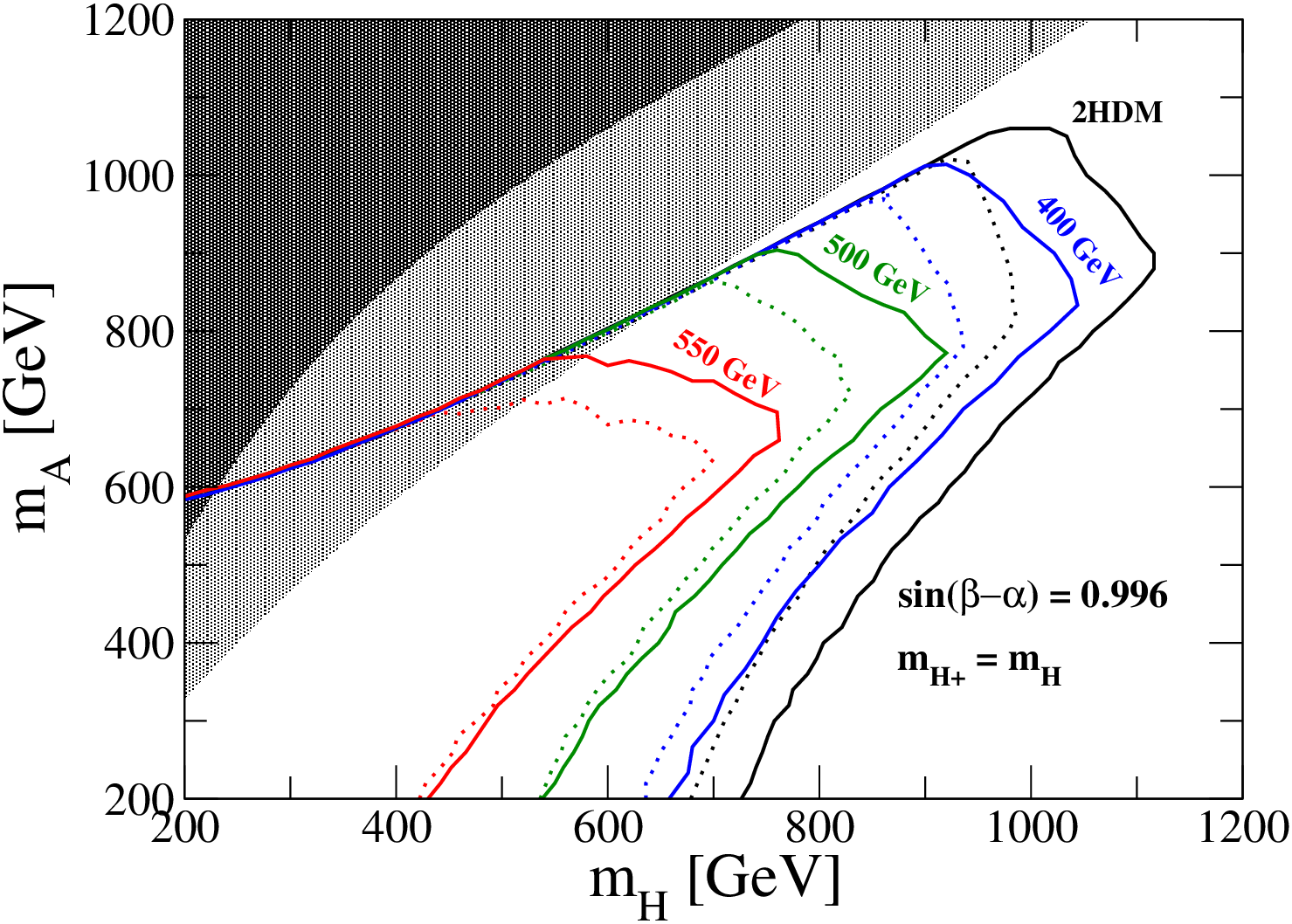}
\caption{Same as Fig.~\ref{const1}, but we take $m_{H^\pm}=m_H$ instead of $m_{H^\pm}=m_A$. 
}
\label{const2}
\end{center}
\end{figure}

\begin{figure}[t]
\begin{center}
\includegraphics[scale=0.3]{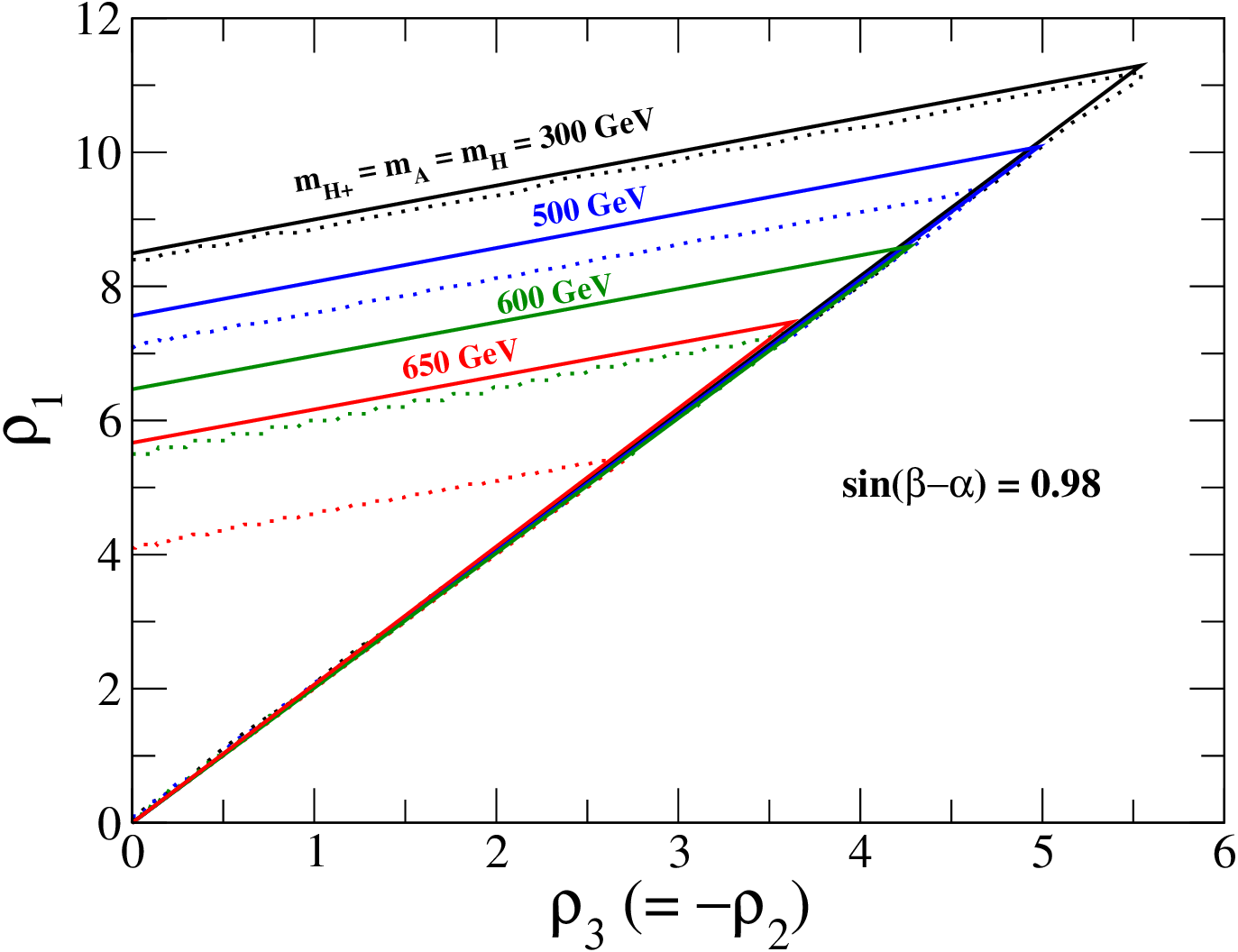} \hspace{3mm}
\includegraphics[scale=0.3]{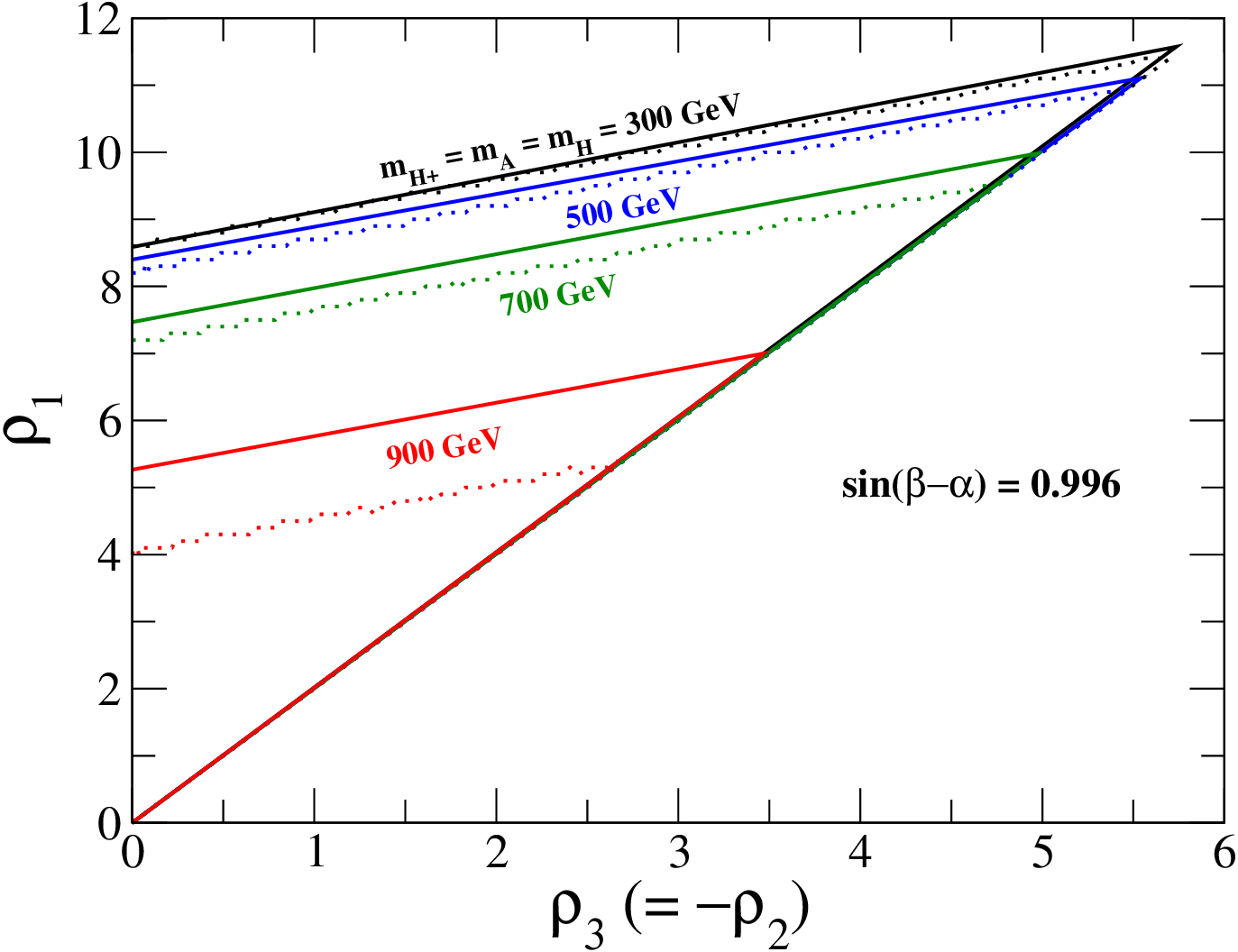}
\caption{
Constraints on the I(1+2)HDM parameter space expressed in the $\rho_3$-$\rho_1$ plane 
induced by unitarity and vacuum stability in the 
case of $m_{H^\pm}=m_A=m_H(\equiv m_\Phi^{})$, $m_{\eta^\pm}=m_{\eta_H}$,  $m_{\eta_A}=63$ GeV and $\sin(\beta-\alpha)=0.98$ (left panel) 
and 0.996 (right panel).
We take $m_{\Phi}^{}=300$[300], 500[500], 600[700] and 650[900] GeV for the black, blue, green and red contours in the left[right] panel. 
The value of $\tan\beta$ is fixed to be 1 in the dotted contours, while it is scanned over the range of $1\leq \tan\beta \leq 30$ in the solid contours. 
For all the plots, we scan the value of $M^2$ in the range of $M^2=m_H^2 \pm 1$ TeV$^2$. 
Outside regions from the contours are excluded by unitarity and vacuum stability. 
}
\label{const3}
\end{center}
\end{figure}

In this section, we discuss how the dimensionless quartic coupling constants in the scalar potential 
can be constrained from perturbative unitarity, vacuum stability and the  $S$ and $T$ parameters in our 3HDMs. 
Among these coupling constants, $\lambda_0$ and $\lambda_{1,...5}$ 
just describe four-point interactions among the inert (pseudo)scalar bosons in the I(1+2)HDM and I(2+1)HDM, respectively, and 
they do not affect to the active sector or the interaction between active and inert sectors at the tree level.  
We thus take these coupling constants to be zero in this section. 
We first show numerical results in the I(1+2)HDM and then in the I(2+1)HDM. 

\subsection{I(1+2)HDM}

In the I(1+2)HDM,  
it is interesting to see how the constraints on 
the parameter space of the masses of active scalar bosons, $H^\pm$, $A$ and $H$,  
can be modified by the requirement of unitarity and vacuum stability as compared to the 2HDM case. 
As we can see in the sections III-A and III-B, the additional coupling constants which do not enter
 in the 2HDM, such as $\rho_i$ and $\sigma_i$, 
appear in the conditions of vacuum stability and unitarity. 
Therefore, a non-zero value of these coupling constants can modify the constraints on the parameters describing the active
(2HDM-like) sector i.e., $\lambda_1$-$\lambda_5$, which 
are translated into bounds on the masses of active (pseudo)scalar bosons. 

In the following, we fix the mass of either $\eta_A$ or $\eta_H$ to be $m_h/2=63$ GeV in order to satisfy the relic abundance of DM~\cite{IDM_DM}, and
take all the other masses of the inert (pseudo)scalar bosons to be larger than $m_h/2$. 
For definiteness, we choose $\eta_A^{}$ as the DM candidate, so that we take $m_{\eta_A^{}}=m_h/2$. 
To avoid a large contribution to the $T$ parameter, we take $m_{H^\pm}=m_A$ or $m_{H^\pm}=m_H$ and $m_{\eta^\pm}=m_{\eta_H}$.  
For simplicity, we take $\sigma_i =\rho_i$ ($i=1,...3$). 
In that case, the coupling constants $\rho_{1,...3}$ are expressed in terms of the masses of inert scalar bosons as
follows:
\begin{align}
\rho_1 &= \frac{2}{v^2}(m_{\eta^\pm}^2-\mu_0^2) , \\
\rho_2 &= \frac{1}{v^2}(m_{\eta_H}^2+m_{\eta_A}^2-2m_{\eta^\pm}^2)=\frac{1}{v^2}(m_{\eta_A}^2-m_{\eta_H^{}}^2) , \\
\rho_3 &= \frac{1}{v^2}(m_{\eta_H}^2-m_{\eta_A}^2)=-\rho_2. 
\end{align}
We can see that the sign of $\rho_3$ is now positive due to the assumption of $m_{\eta_A^{}}=m_h/2 ~(< m_{\eta_H^{}})$. 
The sign becomes negative when we consider the other case, of $\eta_H^{}$ as the DM candidate, but 
the bounds from vacuum stability and unitarity do not depend on the sign of $\rho_3$. 
From the vacuum stability condition of (\ref{vs2}), the value of $\rho_1$ should be larger than $2|\rho_2|$, thus we take  
$\rho_1 \gtrsim 2|\rho_2|$.  

As we discussed in Sec.~II-B, when $\sin(\beta-\alpha)\neq 1$ is taken, we cannot have the decoupling limit of the extra active Higgs bosons and thus we can obtain an upper limit of their masses. 
In addition, recall that the value of $\sin(\beta-\alpha)$ describes the deviation in the $hVV$ couplings as shown in Eq.~(\ref{couplings}) at the tree level so that 
we may be able to find a relationship between the upper limit on the masses and the deviation in the $hVV$ couplings. 
Because the 1-$\sigma$ error of the measurement of $hVV$ couplings is expected to be 2\% at the HL-LHC and 0.4\% at the ILC500~\cite{hVV}, 
we take $\sin(\beta-\alpha)=0.98$ and 0.996, respectively, as examples. 


In Fig.~\ref{const1}, we show the constraint on the parameter space of the I(1+2)HDM mapped
onto the $m_H$-$m_A$ plane by unitarity, vacuum stability and the  $S$ and $T$ parameters. 
We take $\sin(\beta-\alpha)=0.98$ in the left panel and 0.996 in the right panel.  
The value of $\tan\beta$ is fixed to be 1 in the dotted contours, 
while it is scanned over the range  $1\leq \tan\beta \leq 30$ in the solid contours. 
For all the plots, we also scan the value of $M^2$ in the range of $M^2=m_{H}^2\pm 1~\text{TeV}^2$.  
We check that these scanned ranges are enough wide to obtain the maximally allowed region  on the parameter space from the unitarity and vacuum stability constraints. 
The blue, green and red contours show the cases of $m_{\eta^\pm}=400$, 500 and 550 GeV, respectively.  
We also add a plot for the case in the 2HDM as the black contour for comparison. 
The regions outside each contour are excluded by unitarity and vacuum stability. 
In addition, the light and dark shaded regions are excluded by the $S$ and $T$ parameters in the 2HDM and 
 I(1+2)HDM, respectively. 
We note that the contribution to $\Delta T$ is given to be zero due to $m_{H^\pm}=m_A$, 
so that the excluded region denoted by the light and dark area are from the constraint by $\Delta S$. 
Besides, we confirm that the difference in $\Delta S$ due to the different choices of $m_{\eta^\pm}$ is negligible.  

In Fig.~\ref{const2}, we show the case of $m_{H^\pm}=m_H$ with the rest of the setup being the same as in Fig.~\ref{const1}. 
In this case, there is a non-zero contribution to $\Delta T$ which is proportional to $\cos^2(\beta-\alpha)$, so that 
the excluded regions denoted by light and dark shading are different from those in Fig.~\ref{const1}. 

We find that the upper limits on $m_A$ and $m_H$ are related to each other and are getting stronger 
by taking larger values of $m_{\eta^\pm}$. 
We note that, if we take the limit $m_{\eta^\pm}(=m_{\eta_H^{}}) \to m_{\eta_A^{}}$, the constraint from unitarity and vacuum stability approaches to that in the 2HDM. 

In Fig.~\ref{const3}, we show the constraints on the parameter space on the $\rho_3$-$\rho_1$ plane by unitarity and vacuum stability 
in the cases of $\sin(\beta-\alpha)=0.98$ in the left panel and $\sin(\beta-\alpha)=0.996$ in the right panel, where 
the regions outside each contour are excluded. 
The values of $\tan\beta$ and $M^2$ are taken as done in Figs.~\ref{const1} and \ref{const2}. 
The black, blue, green and red contours show the cases of 
$m_{\Phi}^{}=300$[300, 500[500], 600[700] and 650[900] GeV in the left[right] panel, where $m_\Phi^{}=m_{H^\pm}=m_A=m_H$. 
We note that the parameter space shown in this plot is allowed from the $S$ and $T$ parameters. 
We can see that the region with $\rho_1<2\rho_3$ is excluded by the vacuum stability condition in Eq.~(\ref{vs2}). 
Furthermore, the region above each contour is excluded by the unitarity constraints through the largest eigenvalue of the matrix $X_1$ given in Eq.~(\ref{3times3}). 

\subsection{I(2+1)HDM}

Next, we consider an application of our bounds from unitarity and vacuum stability in the I(2+1)HDM. 
As mentioned in the beginning of this section, we take $\lambda_1$-$\lambda_5$ to be zero. 
In that case, we obtain the analytic expression of the eigenvalues for three sub-matrices of the $s$-wave amplitude given in Eq.~(\ref{3times3}) as 
\begin{align}
O_1^TX_1O_1& = \text{diag}\left(0,\frac{3\lambda_0}{2} \pm \sqrt{\frac{9}{4}\lambda_0^2 +(2\rho_1+\rho_2)^2 + (2\sigma_1+\sigma_2)^2}\right) , \label{pu1}\\
O_2^TX_2O_2& = \text{diag}\left(0,\frac{\lambda_0}{2} \pm \sqrt{\frac{1}{4}\lambda_0^2 + \rho_2^2 + \sigma_2^2}\right) , \\
O_3^TX_3O_3& = \text{diag}\left(0,\frac{\lambda_0}{2} \pm \sqrt{\frac{1}{4}\lambda_0^2 + \rho_3^2 + \sigma_3^2}\right),  
\end{align}
where $O_{1,...3}$ are the orthogonal matrices to diagonalize $X_{1,...3}$. Now, $\lambda_0$ is determined by $\lambda_0 = m_h^2/(2v^2)\simeq +0.13$ from Eq.~(\ref{mh}), 
so that the positive eigenvalues are important to get a stronger bound. 

In the following calculation, 
$\eta_{A_1}$ is taken to be the DM candidate with mass $m_h/2$. 
For simplicity, we consider a no-mixing case between two inert scalar states, which can be realized by setting $\mu_{12}^2\simeq 0$\footnote{
When we take $\mu_{12}^2$ to be exactly zero, the lightest state among $\eta_{A_2}^{}$, $\eta_{H_2}^{}$ and $\eta_2^\pm$ is also stable. 
In order to have a decay mode of it, a non-zero value of $\mu_{12}^2$ is required.  }. 
In addition, to avoid the constraint from the $T$ parameter, 
we take $m_{\eta_{H_1}^{}}=m_{\eta_1^\pm}$ and $m_{\eta_{H_2}^{}}=m_{\eta_2^\pm}$. 
We note that results do not change if we replace both[either] $m_{\eta_{H_1}^{}}$ and[or] $m_{\eta_{H_2}^{}}$ with $m_{\eta_{A_1}^{}}$ and[or] $m_{\eta_{A_2}^{}}$. 
In the above set up, we obtain the following relations for the quartic coupling constants: 
\begin{align}
\rho_2 = -|\rho_3|, \quad \sigma_2 = -|\sigma_3|. 
\end{align}
From the vacuum stability condition in Eqs.~(\ref{vs2}) and (\ref{vs3}), we have 
\begin{align}
\rho_1 > 2|\rho_2|,\quad \sigma_1 > 2|\sigma_2|. 
\end{align}
By the combination of the bounds from unitarity and vacuum stability, 
the case with $\rho_1 \gtrsim 2|\rho_2|$ and $\sigma_1 \gtrsim 2|\sigma_2|$ gives the largest allowed parameter space. 
In this setup, we find that the eigenvalue of Eq.~(\ref{pu1}) gives the strongest bound. 

\begin{figure}[t]
\begin{center}
\includegraphics[scale=0.3]{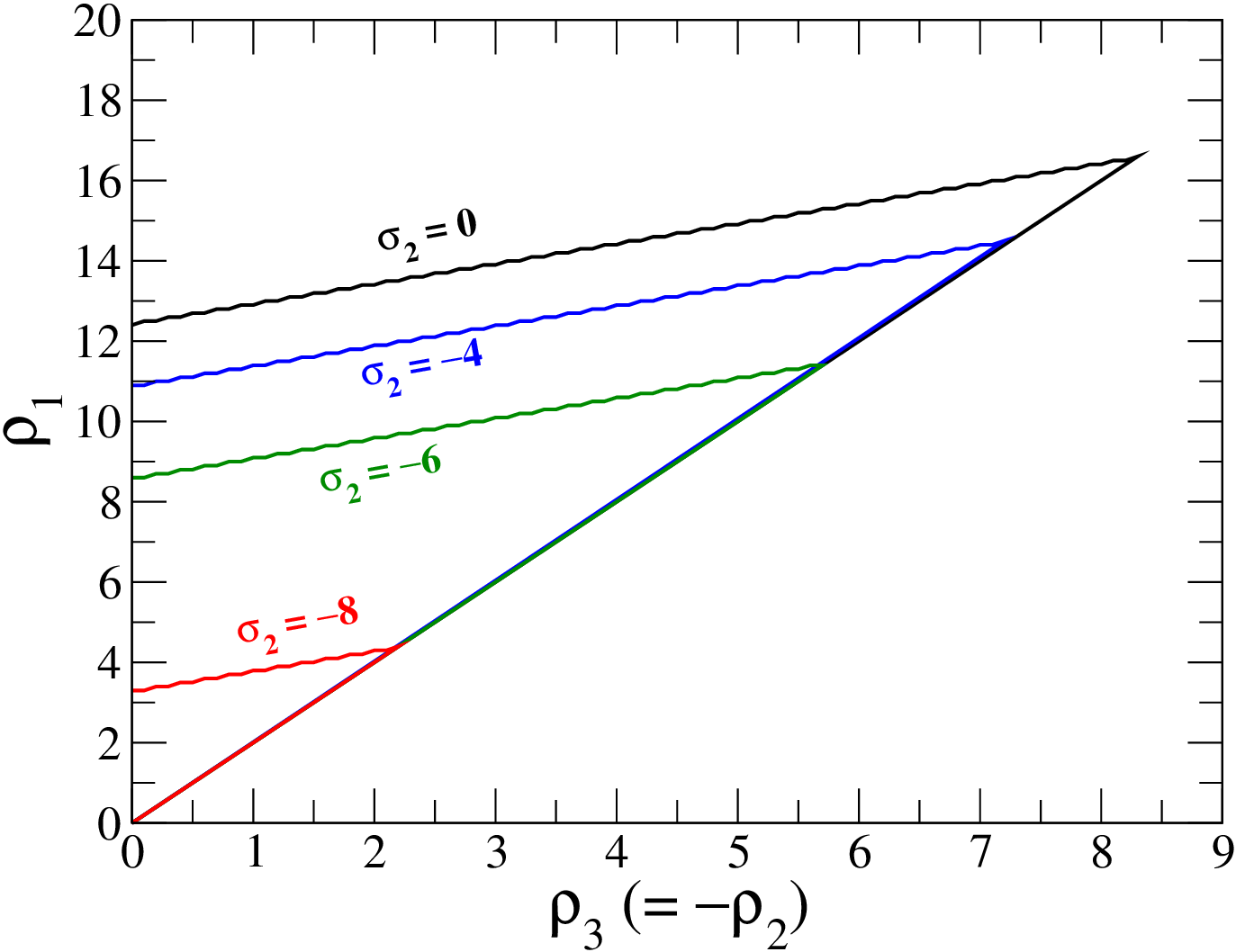} 
\caption{Constraints on the I(2+1)HDM parameter space expressed in the $\rho_3$-$\rho_1$ plane induced by unitarity and vacuum stability.
Outside regions from the contours are excluded by unitarity and vacuum stability. 
}
\label{const4}
\end{center}
\end{figure}

In Fig.~\ref{const4}, we show the allowed parameter space mapped onto the $\rho_3$-$\rho_1$ plane in the I(2+1)HDM.
Regions outside the contours are excluded by unitarity and vacuum stability. 
More precisely, the region below the contour with $\rho_1 = 2\rho_3$ is excluded by vacuum stability and 
the region above the contour is excluded by unitarity. 

\section{Conclusions}

We have discussed the constraints on the parameter space of the I(1+2)HDM and I(2+1)HDM 
emerging from the bounds on vacuum stability, perturbative unitarity and the EW $S$, $T$  and $U$ parameters.  
To impose the unitarity bound, 
we have calculated the $s$-wave amplitude matrix for all possible 2--to--2 body (pseudo)scalar boson elastic scatterings in the high energy limit by means of the ETA, 
where the sub-matrices with $30\times 30$, $36\times 36$ and $12\times 12$ forms for the electromagnetically neutral, singly-charged and doubly-charged channels, respectively, 
can be in block diagonal form with $3\times 3$ sub-matrices as the largest component. 
We have first applied our formulae for the unitarity and vacuum stability bounds to constrain the parameter space in the I(1+2)HDM, especially mapped onto the $m_H$-$m_A$ plane.  
We have found that  
larger excluded regions on such a plane can be obtained in the case with $\sin(\beta-\alpha)\neq 1$ 
as compared to those in 2HDMs. We then have applied our procedure to the I(2+1)HDM on the $\rho_3$-$\rho_1$ plane. 
Here, it has been clarified that the values $\rho_3\gtrsim 8$, 7, 6 and 2 are respectively excluded for any values of $\rho_1$ in the case of $\sigma_2=0$, $-4$, $-6$ and $-8$. 
\\\\
\noindent
$Acknowledgments$

\noindent
S. M. is financed in part through the NExT
Institute. Both authors acknowledge useful conversations with Venus Keus.
K.~Y. is supported by JSPS postdoctoral fellowships for research abroad.


\begin{appendix}

\end{appendix}

\end{document}